\documentclass{cjp}
\usepackage{epsfig}

\usepackage{color}

\usepackage[colorlinks,%
            citecolor=red,linkcolor=blue,hypertex, %
            breaklinks=true]{hyperref}
\begin{document}
\title{The hyperons in the massive neutron star PSR J0348+0432}
\author{Xian-Feng Zhao$^{1,2 \ast}$
\\ \emph{$^{1}$ School of Sciences, Southwest Petroleum University, Chengdu, 610500, China\\
$^{2}$ School of Electronic and Electrical Engineering, Chuzhou University, Chuzhou, 239000, China}
}
\maketitle
\begin{abstract}
Whether the massive neutron star PSR J0348+0432 can become a hyperon star is examined in the framework of the relativistic mean field theory by adjusting the hyperon coupling constants. It is found that at the central baryon number density of the massive neutron star PSR J0348+0432, the relative particle number density of hyperons is smaller than those of neutrons and so it can not change into a hyperon star. In its center, it is mainly composed of $\Lambda$, $\Xi^{-}$ and a few $\Xi^{0}$. We also found that for the neutron star with a maximum mass of 1.4 M$_{\odot}$, it can change into a hyperon star and the hyperon star transition density is 0.668 fm$^{-3}$, at which the hyperons are only composed of $\Lambda$. At its center, the hyperons are also only composed of $\Lambda$ and the ratio of the hyperons is a little larger than that of the neutrons. This illustrates that the NS1.4M$_{\odot}$ has just changed into a hyperon star.
\end{abstract}


\section{Introduction}
One of the major events of astronomy in 2013 is the discovery of the massive neutron star PSR J0348+0432, which is a $2.01\pm0.04$ M$_{\odot}$ pulsar in a 2.46-hour orbit with a $0.172\pm0.003$ M$_{\odot}$ white dlwarf~\cite{Antoniadis 2013} and may be the largest mass neutron star up to now. Neutron star mass is related to gamma-ray bursts, the emission of gravitational waves, periodic millisecond radio signals and month-long X-ray outbursts~\cite{Ozel 2010}.  Hence, how to theoretically describe the massive mass of the neutron star PSR J0348+0432 would be a very important work for astronomy.

The maximum mass of a neutron star would be affected by the symmetry energy~\cite{Glendenning 1985}, the density dependence of the nucleonic equation of state and the hyperon meson couplings~\cite{Cavagnoli 2011}. In the framework of relativistic mean field (RMF) theory, it has been determined to be in the range of $1.5 \sim 1.97$ M$_{\odot}$ as the baryon octet considered but it can reach 2.36 M$_{\odot}$~\cite{Provid那ncia 2013,Glendenning 1991,Zhaoprc 2012,Bombaci 2008} or about 2.8 M$_{\odot}$ as only nucleons considered~\cite{Fattoyevprc 2010}. But the maximum mass of the neutron star also can arrive at 2.25 M$\odot$ with the hyperons being considered and the suitable couplings being chosen~\cite{Colucci 2013}. The obvious is the hyperons in neutron star can reduce the mass.

Neutron stars are high density objects and their masses depend on internal particles constituents~\cite{Glendenning 1997}. For a neutron star, the more close to the center, the baryon number density becomes greater. With the baryon density increase, the chemical potentials of hyperons increase. When the chemical potential of hyperon is more than its mass, the hyperon produces. The hyperon star is a neutron star, in which the numbers of the hyperons converted from nucleons become greater than those of nucleons.~\cite{Glendenning 1985} and the baryon number density is called the hyperon star transition density~\cite{Zhaocpc 2010}. Whether the massive neutron star PSR J0348+0432 can become a hyperon star is a very interesting question for us.

The neutron star mass is very sensitive to the nucleon coupling constants~\cite{Zhaoass 2011}, one of which is GL85~\cite{Glendenning 1985}. Calculations showed GL85 set can better describe the properties of the neutron star. The neutron star mass is also very sensitive to the hyperon coupling constants~\cite{Zhaoprc 2012}. In practical calculations, the hyperon coupling constants can be chosen by the constituent quark model [SU(6) symmetry] or the coupling constants of mesons $\rho, \omega$ chosen by SU(6) symmetry while those of mesons $\sigma$ chosen by fitting to the $\Lambda, \Sigma$ and $\Xi$ well depth in nuclear matter. The results indicate that the ratio of hyperon coupling constant to nucleon coupling constant is in the range of $\sim$ 1/3 to 1~\cite{Glendenning 1991}.

It is thus obvious that, considering the hyperons in neutron star, the mass of the massive neutron star PSR J0348+0432 may be obtained as we choose the suitable coupling constants.

In this paper, the RMF theory is used to study whether the massive neutron star PSR J0348+0432 can become a hyperon star.

\section{The RMF theory and the mass of a neutron star}
The Lagrangian density of hadron matter reads as follows~\cite{Glendenning 1997}

\begin{eqnarray}
\mathcal{L}&=&
\sum_{B}\overline{\Psi}_{B}(i\gamma_{\mu}\partial^{\mu}-{m}_{B}+g_{\sigma B}\sigma-g_{\omega B}\gamma_{\mu}\omega^{\mu}
\nonumber\\
&&-\frac{1}{2}g_{\rho B}\gamma_{\mu}\tau\cdot\rho^{\mu})\Psi_{B}+\frac{1}{2}\left(\partial_{\mu}\sigma\partial^{\mu}\sigma-m_{\sigma}^{2}\sigma^{2}\right)
\nonumber\\
&&-\frac{1}{4}\omega_{\mu \nu}\omega^{\mu \nu}+\frac{1}{2}m_{\omega}^{2}\omega_{\mu}\omega^{\mu}-\frac{1}{4}\rho_{\mu \nu}\cdot\rho^{\mu \nu}
\nonumber\\
&&+\frac{1}{2}m_{\rho}^{2}\rho_{\mu}\cdot\rho^\mu-\frac{1}{3}g_{2}\sigma^{3}-\frac{1}{4}g_{3}\sigma^{4}
\nonumber\\
&&+\sum_{\lambda=e,\mu}\overline{\Psi}_{\lambda}\left(i\gamma_{\mu}\partial^{\mu}
-m_{\lambda}\right)\Psi_{\lambda}
.\
\end{eqnarray}
Here, the units $c=G=\hbar=1$ are used.

The energy density and pressure of a neutron star are given by

\begin{eqnarray}
\varepsilon&=&
\frac{1}{2}m_{\sigma}^{2}\sigma^{2}+\frac{1}{3}g_{2}\sigma^{3}+\frac{1}{4}g_{3}
\sigma^{4}+\frac{1}{2}m_{\omega}^{2}\omega_{0}^{2}+\frac{1}{2}m_{\rho}^{2}\rho_{03}^{2}
\nonumber\\
&&+\sum_{B}\frac{2J_{B}+1}{2\pi^{2}}\int_{0}^{\kappa_{B}}\kappa^{2}\mathrm d\kappa\sqrt{\kappa^{2}+m_{{\color{blue}B}}^{*2}}
\nonumber\\
&&+\frac{1}{3}\sum_{\lambda=e,\mu}\frac{1}{\pi^{2}}\int_{0}^{\kappa_{\lambda}}\kappa^{2}\mathrm d\kappa\sqrt{\kappa^{2}+m_{\lambda}^{*2}}
,\\
p&=&
-\frac{1}{2}m_{\sigma}^{2}\sigma^{2}-\frac{1}{3}g_{2}\sigma^{3}-\frac{1}{4}g_{3}
\sigma^{4}+\frac{1}{2}m_{\omega}^{2}\omega_{0}^{2}+\frac{1}{2}m_{\rho}^{2}\rho_{03}^{2}
\nonumber\\
&&+\frac{1}{3}\sum_{B}\frac{2J_{B}+1}{2\pi^{2}}\int_{0}^{\kappa_{B}}\frac{\kappa^{4}}{\sqrt{\kappa^{2}+m_{B}^{*2}}}\mathrm d\kappa
\nonumber\\
&&+\frac{1}{3}\sum_{\lambda=e,\mu}\frac{1}{\pi^{2}}\int_{0}^{\kappa_{\lambda}}\frac{\kappa^{4}}{\sqrt{\kappa^{2}+m_{\lambda}^{*2}}}\mathrm d\kappa
.\
\end{eqnarray}
where, $m_{B}^{*}$ is the effective mass of baryons
\begin{eqnarray}
m_{B}^{*}=m_{B}-g_{\sigma B}\sigma.
\end{eqnarray}

We use the Oppenheimer-Volkoff ( O-V ) equation to obtain the mass and the radius of {\color{blue}a} neutron star
\begin{eqnarray}
\frac{\mathrm dp}{\mathrm dr}&=&-\frac{\left(p+\varepsilon\right)\left(M+4\pi r^{3}p\right)}{r \left(r-2M \right)}
,\\\
M&=&4\pi\int_{0}^{r}\varepsilon r^{2}\mathrm dr
.\
\end{eqnarray}

\section{Parameters}
In our past work~\cite{Zhaoappb 2012}, the mass of the neutron star was calculated only considering nucleons in it. Corresponding to the various nucleon coupling constants, such as CZ11~\cite{Zhaoijtp2011}, DD$-$MEI~\cite{Typel1999}, GL85~\cite{Glendenning 1985}, GL97~\cite{Glendenning 1997}, NL1~\cite{Suk-Joon1986}, NL2~\cite{Suk-Joon1986}, NLSH~\cite{Sharma1993}, TM1~\cite{Sugahara1994} and TM2~\cite{Sugahara1994}, the mass{\color{blue}es} of the neutron star{\color{blue}s} obtained by us are 1.9702 M$_{\odot}$, 2.8152 M$_{\odot}$, 2.1432 M$_{\odot}$, 2.0177 M$_{\odot}$, 2.8278 M$_{\odot}$, 2.6117 M$_{\odot}$, 3.4189 M$_{\odot}$, 2.5777 M$_{\odot}$ and 2.6695 M$_{\odot}$, respectively.

We know that the neutron star mass will decrease as the hyperons are considered. So, for the parameters CZ11 (1.9702 M$_{\odot}$, no hyperons) and GL97 (2.0177 M$_{\odot}$, no hyperons), the maximum mass of the neutron star calculated with hyperons would become smaller compared with that without considered hyperons and therefore they probably can not give the mass of the massive neutron star of PSR J0348+0432 (2.01 M$_{\odot}$). For the other sets of parameters above, the maximum mass obtained only nucleons considered are much more than the mass of the PSR J0348+0432 (2.01 M$_{\odot}$). And we can suppose that the maximum mass obtained would still be greater than that of the massive neutron star even if hyperons are considered. For the parameters GL85, the maximum mass obtained only nucleons considered is 2.1432 M$_{\odot}$ and is not much more than the mass of the massive neutron star of PSR J0348+0432 (2.01 M$_{\odot}$). Thus, in all the parameters above, to choose GL85 set to fit to the mass of the massive neutron star of PSR J0348+0432 (2.01 M$_{\odot}$) may be suitable.

The recent research on nuclear effective interactions showed a maximum neutron star mass of 1.94 M$_\odot$ ~\cite{Fattoyev 2010}, which corresponds to pure neutron matter and is smaller than the mass of the massive neutron star PSR J0348+0432. But it still can not give the mass of the massive neutron star of PSR J0348+0432 (2.01 M$_{\odot}$).

Therefore, in this work, we choose the nucleon coupling constant GL85 set~\cite{Glendenning 1985}: the saturation density $\rho_{0}$=0.145 fm$^{-3}$, binding energy B/A=15.95 MeV, a compression modulus $K=285$ MeV, charge symmetry coefficient $a_{sym}$=36.8 MeV and the effective mass $m^{*}/m$=0.77.

For the hyperon coupling constant, we define the ratios:
\begin{eqnarray}
x_{\sigma h}&=&\frac{g_{\sigma h}}{g_{\sigma}}
,\\
x_{\omega h}&=&\frac{g_{\omega h}}{g_{\omega}}
,\\
x_{\rho h}&=&\frac{g_{\rho h}}{g_{\rho}}
.\
\end{eqnarray}
Here, $h$ denotes hyperons $\Lambda, \Sigma$ and $\Xi$.

According to SU(6) symmetry, we choose $x_{\rho \Lambda}=0, x_{\rho \Sigma}=2, x_{\rho \Xi}=1$~\cite{Schaffner 1996}. The experimental data of the well depth are $U_{\Lambda}^{(N)}=-30$ MeV~\cite{Batty 1997}, $ U_{\Sigma}^{(N)}=10\sim40$ MeV~\cite{Kohno 2006,Harada 2005,Harada 2006,Friedman 2007} and $U_{\Xi}^{(N)}=-18$ MeV~\cite{Schaffner-Bielich 2000}. Here, we choose $U_{\Lambda}^{(N)}=-30$ MeV, $ U_{\Sigma}^{(N)}$=40 MeV and $U_{\Xi}^{(N)}=-18$ MeV. The relationship between the $x_{\sigma h}$ and $x_{\omega h}$ are as fowllows~\cite{Glendenning 1997}

\begin{eqnarray}
U_{h}^{(N)}=m_n\left(\frac{m_{n}^{*}}{m_{n}}-1\right)x_{\sigma h}+\left(\frac{g_{\omega N}}{m_{\omega}}\right)^{2}\rho_{0}x_{\omega h}
.\
\end{eqnarray}
Here, $m_{n}$ and $m^{*}_{n}$ are the nucleon mass and the effective nucleon mass, respectively.

For $x_{\sigma h}$, we choose $x_{\sigma h}$=0.4, 0.5, 0.6, 0.7, 0.8, 0.9,1.0. For each $x_{\sigma h}$, the $x_{\omega h}$ first is chosen as 0.4, 0.5, 0.6, 0.7, 0.8, 0.9,1.0, respectively. Then, with the restriction of the hyperon well depth it will be slightly adjusted. The parameters that fit to the experimental data of the hyperon well depth are listed in Table~\ref{tab1}.

\begin{table}[!htbp]
\begin{center}
\centering
\caption{The hyperon coupling constants fitted to the experimental data of the well depth, which are $U_{\Lambda}^{(N)}=-30$ MeV, $U_{\Sigma}^{(N)}=+40$ MeV and $U_{\Xi}^{(N)}=-18$ MeV, respectively.}
\label{tab1}
\begin{tabular}[t]{llllll}
\hline\noalign{\smallskip}
$x_{\sigma \Lambda}$ &$x_{\omega \Lambda}$&$x_{\sigma \Sigma}$  &$x_{\omega \Sigma}$ &$x_{\sigma \Xi}$     &$x_{\omega \Xi}$    \\
\hline
0.4    &0.3679     &0.4               &0.825             &0.4               &0.4464     \\
0.5    &0.5090     &\underline{0.5}   &\underline{0.966} &0.5               &0.5874     \\
0.6    &0.6500     &                  &                  &0.6               &0.7284     \\
0.7    &0.7909     &                  &                  &0.7               &0.8693     \\
0.8    &0.9319     &                  &                  &                  &                         \\
\noalign{\smallskip}\hline\noalign{\smallskip}
\end{tabular}
\vspace*{1cm}  
\end{center}
\end{table}

Because the positive well depth $U_{\Sigma}^{(N)}$ will restrain the production of the hyperons $\Sigma$~\cite{Zhaoass 2011}, the values of $x_{\sigma \Sigma}, x_{\omega \Sigma}$ almost can not influence the mass of the neutron star. So, we only select $x_{\sigma \Sigma}=0.4, x_{\omega \Sigma}=0.825$ while $x_{\sigma \Sigma}=0.5, x_{\omega \Sigma}=0.966$ is not necessary to be considered. Thus, sequently selecting each set of $x_{\sigma \Lambda} x_{\omega \Lambda}$, $x_{\sigma \Sigma}=0.4, x_{\omega \Sigma}=0.825$ and $x_{\sigma \Xi} x_{\omega \Xi}$ from Table~\ref{tab1}, we can make up 20 sets of suitable parameters ( named as NO.01, NO.02, ..., NO.20 ).

\begin{figure}[!htbp]
\begin{center}
\includegraphics[width=1\columnwidth]{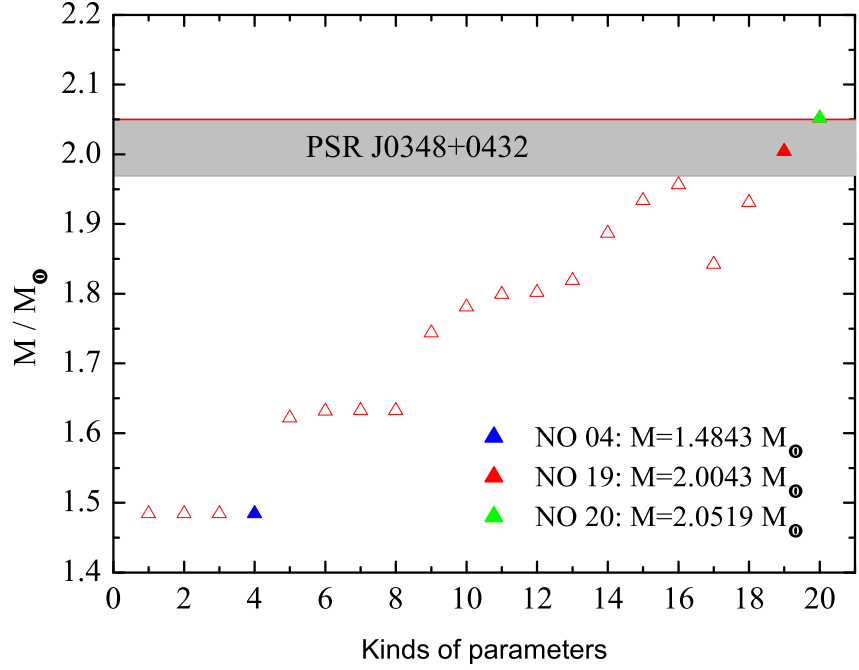}
\caption{The maximum mass of the neutron stars calculated in this work.}
\label{fig1}
\end{center}
\end{figure}

For every set of parameters we calculate the mass of the neutron star. We see that only parameters NO.19 ( $x_{\sigma \Lambda}=0.8, x_{\omega \Lambda}$=0.9319; $x_{\sigma \Sigma}=0.4, x_{\omega \Sigma}=0.825$; $x_{\sigma \Xi}=0.6, x_{\omega \Xi}=0.7284$ ) and NO.20 ( $x_{\sigma \Lambda}=0.8, x_{\omega \Lambda}$=0.9319; $x_{\sigma \Sigma}=0.4, x_{\omega \Sigma}=0.825$; $x_{\sigma \Xi}=0.7, x_{\omega \Xi}=0.8693$ ) can give the mass of the massive neutron star PSR J0348+0432(see Fig.~\ref{fig1}). The central baryon density for each model calculated in this work are listed in Table~\ref{tab2}. Next, we will use the parameters NO.19 and NO.20 to describe the massive neutron star PSR J0348+0432. In addition, for the parameters NO.01 to NO.04, the constants ( $x_{\sigma \Lambda}$=0.4, $x_{\omega \Lambda}$=0.3679; $x_{\sigma \Sigma}$=0.4, $x_{\omega \Sigma}$=0.825 ) are the same but the constants $x_{\sigma \Xi}$ and $x_{\omega \Xi}$ respectively are ( $x_{\sigma \Xi}=0.4$, $x_{\omega \Xi}=0.4464$ ),( $x_{\sigma \Xi}=0.5$, $x_{\omega \Xi}=0.5874$ ),( $x_{\sigma \Xi}=0.6$, $x_{\omega \Xi}=0.7284$ ),( $x_{\sigma \Xi}=0.7$, $x_{\omega \Xi}=0.8693$ ). The four sets of parameters all give the same neutron star mass 1.4843 M$_{\odot}$, which is referred as NS1.4M$_\odot$, and we choose it as a comparison.

\begin{table}[!htbp]
\begin{center}
\centering
\caption{The central baryon density $\rho_{c}$ and the corresponding hyperon star transition density $\rho_{OH}$ for each model calculated in this work.}
\label{tab2}
\begin{tabular}[t]{llllll}
\hline\noalign{\smallskip}
$NO.$  &$\rho_{c}$ &$\rho_{OH}$       &$NO.$             &$\rho_{c}$        &$\rho_{OH}$   \\
       &fm$^{-3}$  &fm$^{-3}$         &                  &fm$^{-3}$         &fm$^{-3}$      \\
\hline
01     &0.698      &0.668             &11                &0.889             &0.761         \\
02     &0.696      &0.668             &12                &0.904             &0.77         \\
03     &0.696      &0.668             &13                &0.812             &0.805          \\
04     &0.696      &0.668             &14                &0.868             &0.819           \\
05     &0.764      &0.683             &15                &0.905             &0.85                         \\
06     &0.8        &0.697             &16                &0.932             &0.895         \\
07     &0.81       &0.699             &17                &0.805             &0.901         \\
08     &0.81       &0.699             &18                &0.856             &0.942          \\
09     &0.805      &0.725             &19                &0.9               &1.002           \\
10     &0.852      &0.739             &20                &0.925             &1.112                         \\
\noalign{\smallskip}\hline\noalign{\smallskip}
\end{tabular}
\vspace*{1cm}  
\end{center}
\end{table}

\begin{figure}[!htbp]
\begin{center}
\includegraphics[width=1\columnwidth]{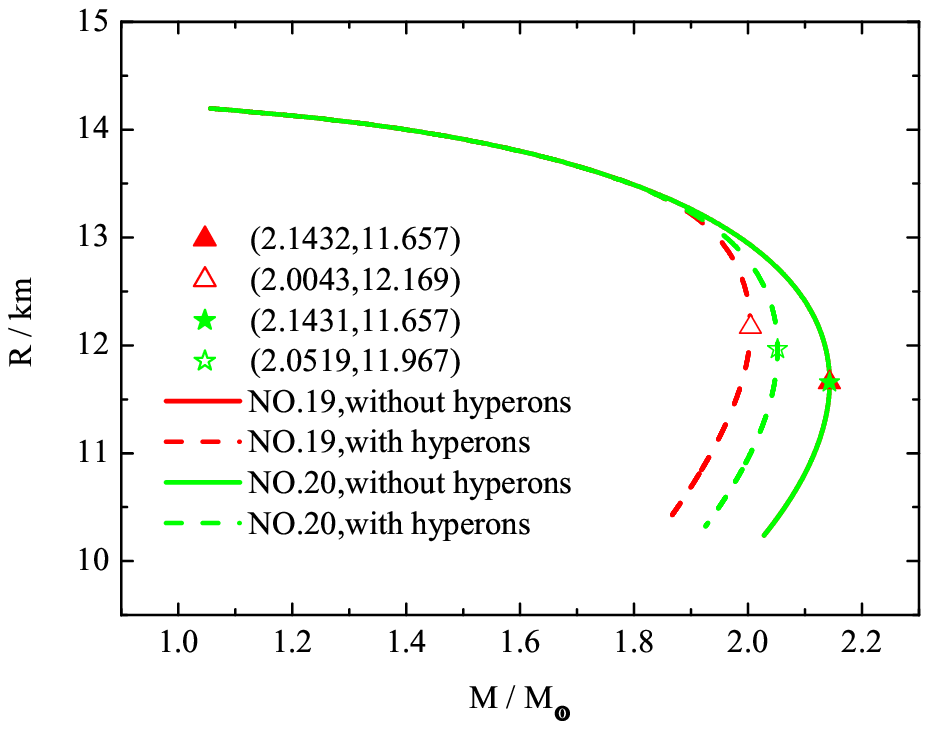}
\caption{The radius as a function of the mass of the neutron stars calculated in this work.}
\label{fig2}
\end{center}
\end{figure}

The radius also can be obtained from eqs.(5) and (6) and hyperons to the neutron star can influence the star＊s radius too. For example, the radius as a function of the mass is shown in Fig.~\ref{fig2} when coupling constant sets 19 and 20 are chosen. We see the mass and the radius all are the same as the hyperons are not considered. When the hyperons are included the radius corresponding to parameters NO.19 is $R$=12.169 km while that corresponding to NO.20 is $R$=11.967 km. The radius given by NO.19 is 0.202 km larger than that by NO.20.

\section{The change from a neutron star to a hyperon star}
\begin{figure}[!htbp]
\begin{center}
\includegraphics[width=1\columnwidth]{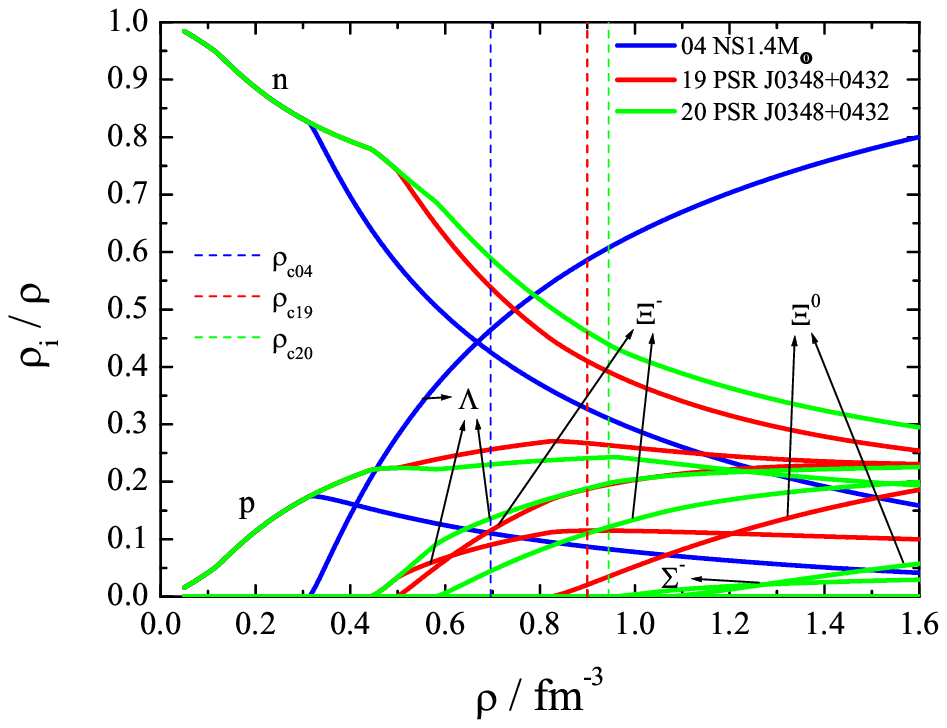}
\caption{The relative particle number density of the baryons as a function of the baryon number density.}
\label{fig3}
\end{center}
\end{figure}

Fig.~\ref{fig3} shows the relative particle number density of the baryons as a function of the baryon number density. In our calculations the baryon density is in the range of $\sim$ 0 to 1.6 fm$^{-3}$. From Table~\ref{tab2} we see that the center baryon number density of the massive neutron star PSR J0348+0432 in our calculation is $\rho_{c19}=0.9$ fm$^{-3}$ (NO.19) or $\rho_{c20}=0.925$ fm$^{-3}$ (NO.20). Therefore, from Fig.~\ref{fig3} we can see that within the massive neutron star PSR J0348+0432, there are five ( n, p, $\Lambda$, $\Xi^{-}$ and $\Xi^{0}$ for NO.19 ) or  four kinds of particles ( n, p, $\Lambda$ and $\Xi^{-}$ for NO.20 ). But for the NS1.4M$_{\odot}$ ( for NO.01 to NO.04 ), whose center baryon number density is $\rho_{c20}=0.696$ fm$^{-3}$, there are only three kinds of particles ( n, p, $\Lambda$ ). Either within the massive neutron star PSR J0348+0432 or within the NS1.4M$_{\odot}$ the hyperons $\Sigma^{-}$, $\Sigma^{0}$ and $\Sigma^{+}$ all don't appear. Moreover, for the NS1.4M$_{\odot}$ the hyperons $\Xi^{-}$ and $\Xi^{0}$ all do not appear.

The hyperon star transition density of a neutron star is defined as the lowest baryon density~\cite{Zhaocpc 2010}

\begin{eqnarray}
\rho_{0H}=\sum_{B} \rho_{B},
\end{eqnarray}
at which
\begin{eqnarray}
\sum_{h} \rho_{h}>\rho_{n},
\end{eqnarray}
where $\rho_{h}$ is the hyperon number density and $\rho_{n}$ is the neutron number density.

\begin{figure}[!htbp]
\begin{center}
\includegraphics[width=1\columnwidth]{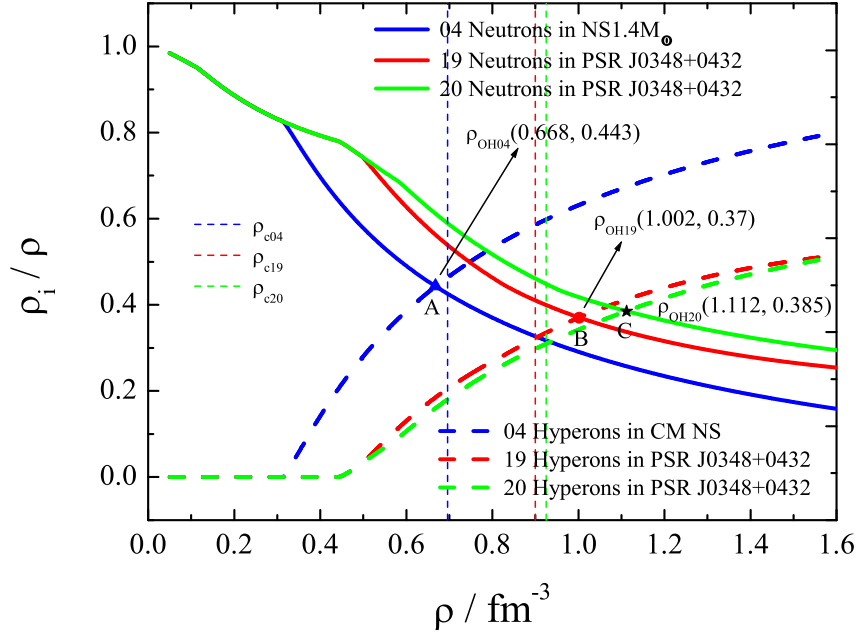}
\caption{The hyperon star transition density of the neutron stars ( for NO.4, NO.19 and No.20) calculated in our work.}
\label{fig4}
\end{center}
\end{figure}

The hyperon star transition density of the neutron star calculated in this work is given in Fig.~\ref{fig4}. We see that the hyperon star transition density of the massive neutron star PSR J0348+0432 is $\rho_{0H19}$=1.002 fm$^{-3}$ ( NO.19 ) or $\rho_{0H20}$=1.112 fm$^{-3}$ ( NO.20 ). The central baryon number density for the massive neutron star PSR J0348+0432 is $\rho_{c19}$=0.9 fm$^{-3}$ ( NO.19 ) or $\rho_{c20}$=0.925 fm$^{-3}$ ( NO.20 ). For the previous two cases, the hyperon star transition density are all larger than the corresponding central baryon number density. Therefore, the massive neutron star PSR J0348+0432 can not change into hyperon star and the hyperon star transition density calculated by us is only theoretical. For parameter NO.04, its hyperon star transition density is $\rho_{0H04}$=0.668 fm$^{-3}$ and its central baryon number density is $\rho_{c04}$=0.696 fm$^{-3}$. Parameters NO.01, NO.02, NO.03 and NO.04 all have the same maximum mass M=1.4843 M $_{\odot}$, which are corresponding to the same radius $R$=13.245 km, respectively. So we can think that the $\rho_{0H04}$ and the $\rho_{c04}$ of NO.01, NO.02 and NO.03 are the same as those of NO.04. Therefore, we can conclude that for NS1.4M$_{\odot}$, its hyperon star transition density and its central baryon number density also are $\rho_{0H04}$=0.668 fm$^{-3}$ and $\rho_{c04}$=0.696 fm$^{-3}$, respectively. Because of $\rho_{0H04}<\rho_{c04}$, the NS1.4M$_{\odot}$ can change into a hyperon star. The mass obtained by parameter NO.5 is $M=$1.6217 M$_{\odot}$, which is corresponding to the radius $R=$12.975 km. For it is different from that obtained by parameters NO.01 to NO.04, we can conclude that the hyperon star transition density of the neutron star calculated by NO.05 must be different from those by parameters NO.01 to NO.04.

\begin{figure}[!htbp]
\begin{center}
\includegraphics[width=1\columnwidth]{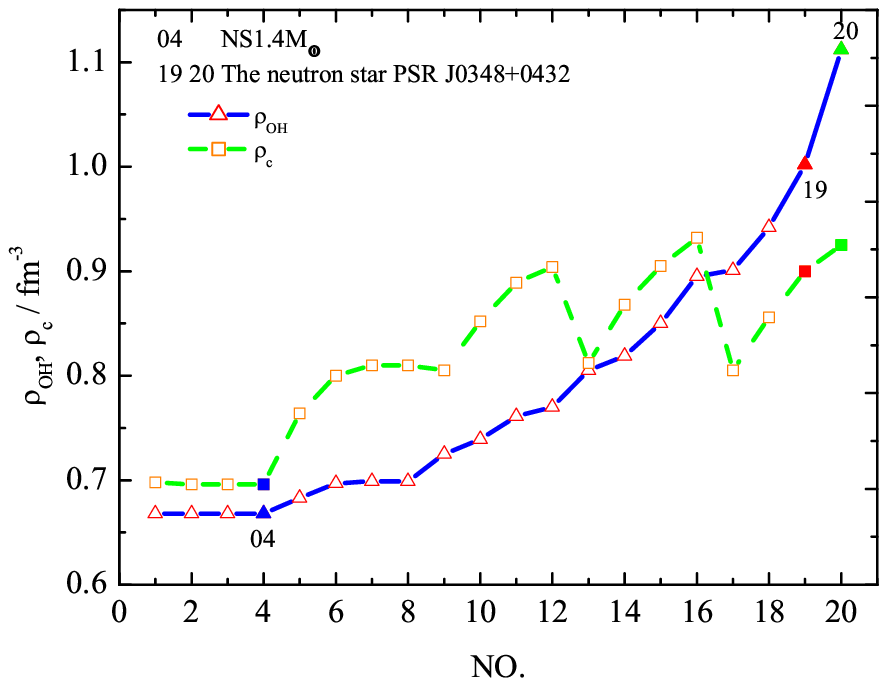}
\caption{The hyperon star transition density of the neutron stars ( for NO.1$\sim$ NO.20) calculated in this work.}
\label{fig5}
\end{center}
\end{figure}

Figure~\ref{fig5} shows the hyperon star transition density of the neutron stars ( for NO.1$\sim$ NO.20) calculated by us. For the cases NO.01$\sim$No.16, the hyperon star transition density $\rho_{OH}$ are all less than the central baryon number density$\rho_{c}$. So for these 16 cases, the neutron stars all can change into hyperon stars at their respective hyperon star transition density $\rho_{OH}$. But for the cases NO.17, NO.18 and the massive neutron star PSR J0348+0432 (NO.19 and NO.20), the hyperon star transition density $\rho_{OH}$ all are greater than the central baryon number density$\rho_{c}$. Therefore, for these 4 cases, the neutron stars can not change into hyperon stars. For real stars, these hyperon star transition density do not exist.

\section{The particle constituent of the neutron stars at their hyperon star transition density and/or their central baryon number density}

The hyperon star transition density calculated by NO.04 is $\rho_{0H}$=0.668 fm$^{-3}$. For the cases of NO.19 and NO.20, the neutron stars can not change into hyperon stars. Even so, we still want to know by which particles the neutron stars are composed at their central baryon number density. Thus, next we will examine the particle constituent of the NS1.4M$_{\odot}$ ( NO.04 ) and the massive neutron star PSR J0348+0432 at the such following baryon densities: $\rho=$0.5 fm$^{-3}$, $\rho=\rho_{OH04}$=0.668 fm$^{-3}$, $\rho=\rho_{c04}$=0.696 fm$^{-3}$, $\rho=\rho_{c19}$=0.9 fm$^{-3}$ and $\rho=\rho_{c20}$=0.925 fm$^{-3}$.

\begin{figure}[!htbp]
\begin{center}
\includegraphics[width=1\columnwidth]{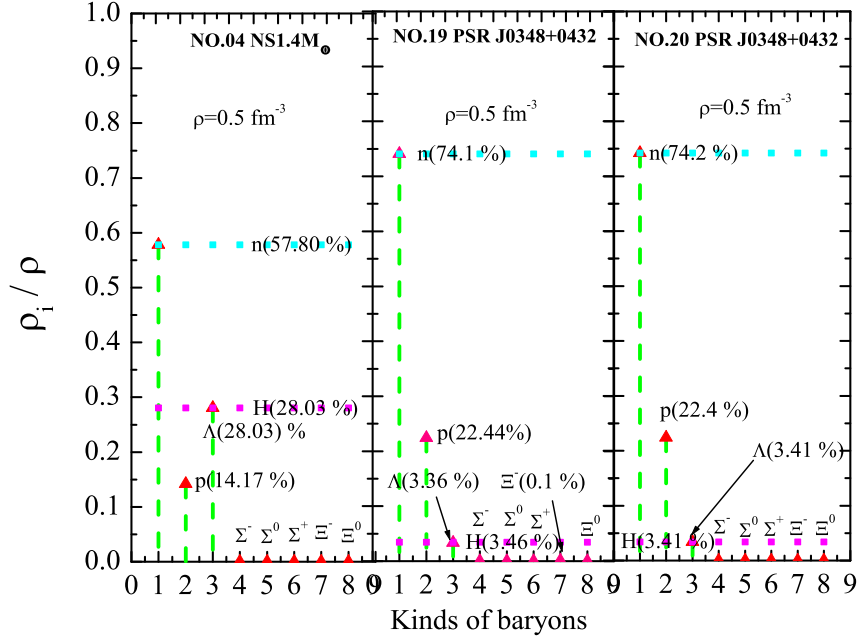}
\caption{The relative particle number density of the baryons as the baryon number density $\rho$=0.5 fm$^{-3}$.}
\label{fig6}
\end{center}
\end{figure}

The relative particle number density of the baryons as the baryon number density $\rho$=0.5 fm$^{-3}$ are shown in Fig.~\ref{fig6}. The abscissa indicates the kinds of baryons: $1-n$, $2-p$, $3-\Lambda$, $4-\Sigma^{-}$, $5-\Sigma^{0}$, $6-\Sigma^{+}$, $7-\Xi^{-}$ and $8-\Xi^{0}$. The ordinate indicates the relative particle number density of each bayon. As the baryon number density $\rho$=0.5 fm$^{-3}$, for the NS1.4M$_{\odot}$ ( NO.04 ) the relative particle number density of neutrons is 57.8 \% while that of hyperons is 28.03 \%. For the massive neutron star PSR J0348+0432, the relative particle number density of neutrons and hyperons are 74.1 \% and 3.46 \% ( NO.19 ) or 74.2 \% and 3.41 \% ( NO.20 ), respectively. For all the three cases, both the massive neutron star PSR J0348+0432 and the NS1.4M$_{\odot}$ are mainly composed of neutrons and they have not yet changed into hyperon stars.

\begin{figure}[!htbp]
\begin{center}
\includegraphics[width=1\columnwidth]{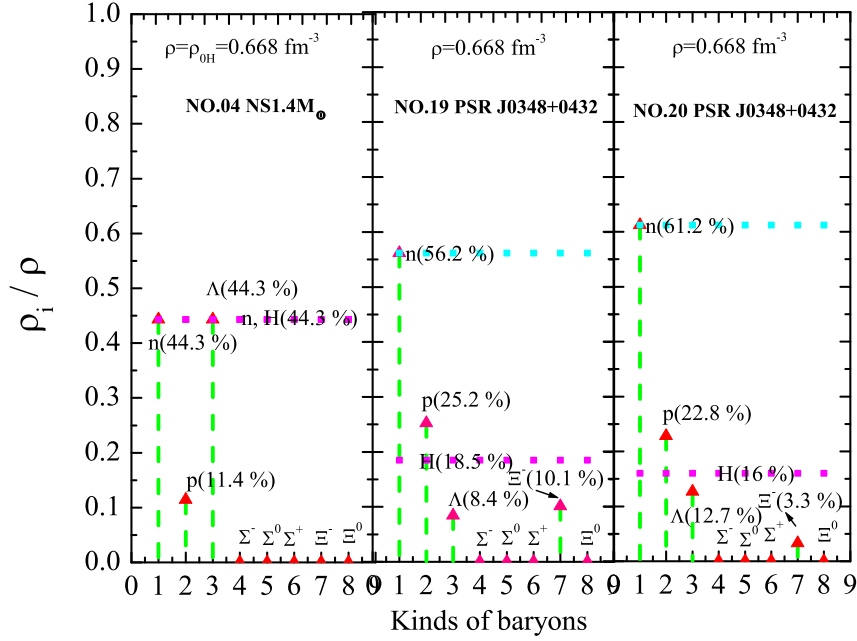}
\caption{The relative particle number density of the baryons as the baryon number density $\rho$=0.668 fm$^{-3}$.}
\label{fig7}
\end{center}
\end{figure}

Fig.~\ref{fig7} gives the relative particle number density of the baryons as the baryon number density $\rho$=0.668 fm$^{-3}$. We see from Fig.~\ref{fig4} that $\rho$=0.668 fm$^{-3}$ is the hyperon star transition density $\rho_{0H04}$ of the NS1.4M$_{\odot}$. As the baryon number density $\rho=\rho_{OH04}$=0.668 fm$^{-3}$, for the NS1.4M$_{\odot}$ ( NO.04 ) the relative particle number density of neutrons and hyperons all are equal to 44.3 \%. Right now, the NS1.4M$_{\odot}$ begins to change into a hyperon star. Here, the hyperons are only composed of $\Lambda$ ( 44.3 \% ). While for the massive neutron star PSR J0348+0432, the relative particle number density of neutrons and hyperons are 56.2 \% and 18.5 \% for NO.19 or 61.2 \% and 16 \% for NO.20, respectively. This means at this moment the massive neutron star PSR J0348+0432 is mainly composed of neutrons and it has not yet changed into a hyperon star.

\begin{figure}[!htbp]
\begin{center}
\includegraphics[width=1\columnwidth]{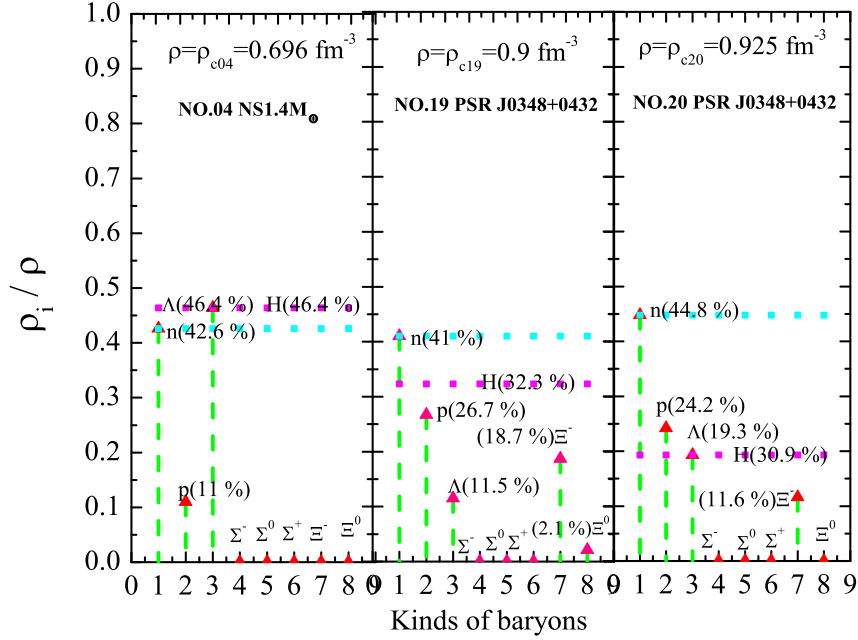}
\caption{The relative particle number density of the baryons at the central baryon number density for the neutron stars calculated by parameters NO.04, NO.19 and NO.20.}
\label{fig8}
\end{center}
\end{figure}

Fig.~\ref{fig8} shows the particle composition at each central baryon number density for the neutron stars calculated by parameters NO.04, NO.19 and NO.20. At the central baryon number density of the NS1.4M$_{\odot}$ ( NO.04, $\rho_{c04}$=0.696 fm$^{-3}$ ), the relative particle number density of neutrons and hyperons are 42.6 \% and 46.4 \%, respectively. The hyperons are only composed of $\Lambda$ and the ratio of the hyperons is a little larger than that of the neutrons. This illustrates that the NS1.4M$_{\odot}$ has just changed into a hyperon star.

For the massive neutron star PSR J0348+0432 calculated by NO.19, the relative particle number density of neutrons and hyperons in the center ($\rho_{c19}$=0.9 fm$^{-3}$) are 41 \% and 32.3 \%, respectively. In this case, the hyperons are mainly composed of $\Lambda(11.5 \%)$, $\Xi^{-}(18.7 \%)$ and a few $\Xi^{0}(2.1 \%)$. For the massive neutron star PSR J0348+0432 calculated by NO.20, the relative particle number density of the neutrons and the hyperons in the center ($\rho_{c19}$=0.925 fm$^{-3}$) are 44.8 \% and 30.9 \%, respectively. Here, the hyperons are mainly composed of $\Lambda(19.3 \%)$ and $\Xi^{-}(11.6 \%)$ while the hyperons $\Xi^{0}$ dose not appear. For the previous two cases of NO.19 and NO.20, the relative particle number density of neutrons all are greater than those of the hyperons at their central baryon number density. This means that the massive neutron star PSR J0348+0432 can not become a hyperon star.

\section{Summary}
In this paper, the hyperon star transition density $\rho_{0H}$  of the massive neutron star PSR J0348+0432 is examined in the framework of the RMF theory with the hyperon coupling constants chosen by us. We find that, for parameters NO.19 and NO.20, the relative particle number density of neutrons all are greater than those of the hyperons at their central baryon number density and therefore the massive neutron star PSR J0348+0432 can not become a hyperon star. Hyperons at the center of the massive neutron star PSR J0348+0432 are mainly composed of $\Lambda$, $\Xi^-$, and a few $\Xi^0$.

We also see that the NS1.4M$_{\odot}$ can change into a hyperon star. The hyperon star transition density of the NS1.4M$_{\odot}$ is $\rho=\rho_{OH04}$=0.668 fm$^{-3}$, at which the hyperons are only composed of $\Lambda$ ( 44.3 \% ). At the central baryon number density for the NS1.4M$_{\odot}$ ( NO.04, $\rho_{c04}$=0.696 fm$^{-3}$ ), the relative particle number density of neutrons and hyperons are 42.6 \% and 46.4 \%, respectively. The hyperons are only composed of $\Lambda$ and the ratio of the hyperons is a little larger than that of the neutrons. This illustrates that the NS1.4M$_{\odot}$ can change into a hyperon star.

The calculations above show that set NO.04 with neutron star mass about 1.4M$_{\odot}$ can turn to a hyperon star, while sets NO.19 and NO.20 with neutron star masses beyond 2.0M$_{\odot}$ can not. This is because there exists more neutrons than hyperons in the massive neutron star compared with that in the NS1.4M$_{\odot}$.

By now no experimental observations have been found that the hyperon star exists. Our conclusions totally result from the RMF method and our chosen of coupling constants. This method is not the absolute correct theory and there may be rooms for improvements.

\textbf{Acknowledgements}\\
We are thankful to the anonymous referee for many useful comments and suggestions. This work was supported by the Special Funds for Theoretical Physics Research Program of the Natural Science Foundation of China ( Grant No. 11447003 ) and the Scientific Research Foundation of the Higher Education Institutions of Anhui Province, China ( Grant No. KJ2014A182 ).

\end{document}